\documentclass[11pt]{article}

\usepackage{hyperref}
\usepackage{fullpage}
\usepackage{graphicx}
\usepackage{amssymb}
\usepackage{natbib}
\usepackage{amsmath}
\usepackage{graphics}
\usepackage{bm}
\usepackage{color}
\usepackage{amsthm}
\usepackage[plain, linesnumbered, lined, ruled, vlined]{algorithm2e}
\usepackage{bbm}
\usepackage{enumerate}
\usepackage{enumitem}

\newcommand{\E}{\mathbb{E}}

\newcommand{\pr}{\mathrm{Pr}}
\DeclareMathOperator*{\argmin}{arg\,min}
\newcommand{\remind}[1]{}

\newtheorem{proposition}{Proposition}
\newtheorem{assumption}{Assumption}
\newtheorem{theorem}{Theorem}

\newtheorem{ex}{Example}
\newcommand{\bigO}{\ensuremath{\mathop{}\mathopen{}\mathcal{O}\mathopen{}}}
\newcommand{\HK}{\mathcal{H}(K)}
\newcommand{\HKK}{\mathcal{H}(K\otimes K)}

\newcommand{\SK}{\mathcal{S}(K)}
\newcommand{\SpK}{\mathcal{S}^{+}(K)}
\newcommand{\tell}{\tilde{\ell}}

\newcommand{\ra}{q}
\newcommand{\tK}{\tilde{K}}
\newcommand{\ip}[1]{\langle #1 \rangle}
\newcommand{\norm}[1]{\| #1 \|}

\newcommand{\svec}{\mathrm{svec}}
\newcommand{\vareps}{\varepsilon}

\renewcommand{\baselinestretch}{1.81}
\newcommand{\linsp}{\renewcommand{\baselinestretch}{1.81}}
\newcommand{\linsps}{\renewcommand{\baselinestretch}{1}}

\begin{document}

\linsps

{
 \title{\bf Nonparametric Operator-Regularized Covariance Function Estimation for Functional Data}
 \author{
  Raymond K. W. Wong\\
  Department of Statistics, Iowa State University\\
  \\
  Xiaoke Zhang \\
  Department of Applied Economics and Statistics, University of Delaware}
  \date{January 22, 2017}
  \maketitle
}

\bigskip
\begin{abstract}
In functional data analysis (FDA), covariance function is fundamental not only as a critical quantity for understanding elementary aspects of functional data but also as an indispensable ingredient for many advanced FDA methods. This paper develops a new class of nonparametric covariance function estimators in terms of various spectral regularizations of an operator associated with a reproducing kernel Hilbert space.
Despite their nonparametric nature, the covariance estimators are automatically positive semi-definite without any additional modification steps.
An unconventional representer theorem is established to provide a finite dimensional representation for this class of covariance estimators, which leads to a closed-form expression of the corresponding $L^2$ eigen-decomposition. Trace-norm regularization is particularly studied to further achieve a low-rank representation, another desirable property which leads to dimension reduction and is often needed in advanced FDA approaches. An efficient algorithm is developed based on the accelerated proximal gradient method.
This resulted estimator is shown to enjoy an excellent rate of convergence under both fixed and random designs. The outstanding practical performance of the trace-norm-regularized covariance estimator is demonstrated by a simulation study and the analysis of a traffic dataset.
\end{abstract}

\noindent
\def\baselinestretch{1}\selectfont

\linsp\selectfont

\section{Introduction}\label{sec:intro}

In recent decades, functional data analysis (FDA) has received substantial attention and become increasingly important especially as the advent of ``Big Data'' era.
Representative monographs on FDA
include
\cite{Ramsay-Silverman05}, \cite{FerrV06}, \cite{HorvK12}, and \cite{HsinE15}.
Typically functional data are collected from $n$ curves $\{X_i: i=1, \ldots, n\}$ that are regarded as independent copies of a real-valued $L^2$
stochastic process $X$ defined on a compact domain $\mathcal{T}$ with mean function $\mu_0(t) = \E \{X(t)\}, t \in \mathcal{T},$ and covariance function $C_0(s, t) = \text{cov}\{X(s), X(t)\}, s, t \in \mathcal{T}$. In reality, due to discrete recording and the presence of noise, the data are often represented by
$\{(T_{ij},Y_{ij}): i=1,\dots, n; j=1,\dots m_i\}$,
where $m_i$ is the number of observations from the $i$-th curve $X_i$, and $Y_{ij} $ is
the noisy  observation from $X_i$ measured at the discrete time point
$T_{ij}$,
i.e., $Y_{ij} =
X_i(T_{ij}) +\varepsilon_{ij}$. Here $\{\varepsilon_{ij}: i=1,\dots, n;
j=1,\dots m_i\}$ are independent errors with zero mean and finite variance. 
For simplicity and without loss of generality we assume $m_i=m$ for all $i$.

Among various population quantities, the covariance function $C_0$ is fundamental in FDA. Generally $C_0$ has two major roles. It is not only an important quantity that characterizes the temporal dependency \citep{Yao-Muller-Wang05, ZhanC07, Li-Hsing10, ZhanW16},
but also a building block for more advanced approaches in FDA
such as functional principal component analysis (FPCA) and functional linear regression \citep{YaoMW05, Hall-Muller-Wang06}.
Hence the FDA literature that involves covariance function estimation may accordingly be categorized into two types depending on the role of $C_0$.
As for the estimation of $C_0$, a variety of nonparametric methods have been proposed, such as local polynomial smoothing \citep{Li-Hsing10, ZhanW16},  B-splines \citep{JameHS00, RiceW01, Paul-Peng09}, penalized splines \citep{Goldsmith-Bobb-Crainiceanu11, XiaoLR13}, and smoothing splines \citep{Rice-Silverman91, Cai-Yuan10}.

Positive semi-definiteness is an essential characteristic of covariance functions.  Therefore, a valid covariance estimator is usually desired to be positive semi-definite, especially when this estimator is involved in subsequent analyses. See \cite{YaoMW05} and Section~\ref{sec:data} for examples. Meanwhile, it is also appealing if a covariance estimator is of low rank since this will encourage dimension reduction, alleviate computational and storage burdens, and facilitate simple interpretations.
In addition, low rank is often needed in trajectory prediction and some other advanced FDA methods \citep[e.g.,][]{Yao-Muller-Wang05, DelaH12,  LiWC13, JianAW16}.
Unfortunately, a majority of existing methods in FDA cannot directly produce a covariance estimator that is positive semi-definite or of low rank.
Hence a two-step procedure is typically performed in order to achieve at least one property, where a constraint-free covariance estimator is first obtained, then followed by
{a reconstruction step (e.g., via FPCA and truncation).}
See
\cite{Hall-Vial06} and \cite{Poskitt-Sengarapillai13}
for instances.
This two-step procedure, however, is unfavorable since it not only complicates the theoretical analysis of the final estimator, but also makes computation unstable due to the non-smooth truncation.

In this paper, we utilize
a reproducing kernel Hilbert space (RKHS) framework
to achieve a coherent ``one-step"
covariance estimation procedure such that the resulted estimator is automatically both positive semi-definite and of low rank.
The application of RKHS has gained popularity recently in FDA \citep[e.g.,][]{YuanC10, ZhuYZ14, WangR15}. In the same vein as penalized splines \citep[e.g.,][]{Pearce-Wand06} and smoothing splines
\citep[e.g.,][]{Wahba90, EggeL09, Gu13},
we suppose that the sample path of $X$ belongs to a RKHS $\HK$ defined on $\mathcal{T}$, 
with a continuous and square-integrable reproducing kernel $K(\cdot, \cdot)$
defined on $\mathcal{T} \times \mathcal{T}$. A key property of $K$
is the so-called reproducing property:
\[
\langle K(t, \cdot), f(\cdot) \rangle_{\HK} = f(t), \quad \text{for any $t \in \mathcal{T}$ and $f \in \HK$}.
\]
Moreover, $K$ also uniquely determines the inner product and norm of $\mathcal{H}(K)$, denoted by $\langle \cdot, \cdot \rangle_{\HK}$ and $\|\cdot\|_{\HK}$ respectively.
A canonical example of RKHS is the $r$-th order Sobolev-Hilbert space on $\mathcal{T}=[0, 1]$: 
\[
\mathcal{W}^r= \{g: g^{(v)}, v=0, \ldots, r-1, \,\text{are absolutely continuous};\, g^{(r)} \in L^2([0, 1])\},
\]
equipped with the squared norm 
\[
\|g\|^2 = \sum^{r-1}_{v=0} \left\{ \int^1_0 g^{(v)}(t)dt\right\}^2 + \int^1_0 \left\{g^{(r)}(t)\right\}^2 dt.
\] 
In this paper, we use $\mathcal{W}^2$ in all numerical implementations, but establish theoretical results for $\mathcal{W}^r, r \geq 2$, with general equivalent norms.

The RKHS framework was also used by \citet{Cai-Yuan10} for covariance function estimation, which is perhaps the most related
work to ours.
This will be made clear that their estimator is a non-positive semi-definite version of a special case in our general spectral regularization framework.
Under the assumption $\E\|X\|^2_{\mathcal{H}(K)}<\infty$, 
they showed that
$C_0\in\HKK$, where $\mathcal{H}(K \otimes K)$ is the tensor
product RKHS
equipped with the norm $\| \cdot \|_{\HKK}$ and the reproducing kernel
\[
K \otimes K ((s_1, t_1), (s_2, t_2)) = K(s_1, s_2) K(t_1, t_2), \quad s_1,s_2,t_1, t_2\in\mathcal{T}.
\]
This suggests a tensor product RKHS modeling of $C_0$, 
which we also adopt in this paper. With slight abuse of notation, we hereafter also use the notation $\otimes$ to denote the tensor product of functions, i.e., $f\otimes g(s, t) = f(s) g(t)$.

\citet{Cai-Yuan10} proposed
to estimate the covariance function $C_0$ by solving
\begin{equation}
  \underset{C\in\HKK}{\min} \left\{ \ell(C) + \lambda \| C\|^2_{\HKK} \right\},  \label{eqn:CY}
\end{equation}
where $\ell$ is a convex and smooth loss function characterizing the fidelity to the data,
and $\lambda > 0$ is a tuning parameter for the penalty term. Unfortunately, this approach cannot ensure the covariance estimator to be positive semi-definite or of low rank, so the aforementioned two-step procedure must be performed to improve the estimator.

We propose a new class of tensor product RKHS covariance estimators via a variety of spectral regularizations of an operator on $\HK$. The spectral regularizations generalize the penalty in (\ref{eqn:CY}), and can easily enable low-rank modeling, e.g., when the trace-norm penalty is used. The estimation framework respects the semi-positivity structure of covariance functions by imposing a constraint, so the resulted estimator automatically inherits this characteristic. Given any penalty, the covariance estimator is obtained by one step, which can reduce the computational and theoretical complexities of the two-step method. We establish a representer theorem to provide a finite dimensional representation for this class of covariance estimators, which makes the estimation procedure practically computable. Compared with its classical counterparts \citep[e.g.,][]{Wahba90, Cai-Yuan10}, the representer theorem is unconventional due to the semi-positivity constraint and a wide range of regularizations (e.g., trace-norm regularization). As a byproduct of the representer theorem, a closed form of the $L^2$ eigen-decomposition admitted by the covariance estimator can be easily obtained, without any numerical approximations needed in common FPCA approaches.

To promote dimension reduction, we particularly focus on trace-norm regularization
to additionally encourage low-rank estimation.
The corresponding objective function involved in the estimation framework is convex but non-differentiable. {An efficient algorithm is developed for this optimization problem} based on the representer theorem and the accelerated proximal gradient method \citep{Beck-Teboulle09}.
Note that, asymptotically, the use of trace-norm regularization does not rule out the
cases when $C_0$ is of high or infinite rank.
Irrespective of the true rank, our
estimator is consistent
with the optimal convergence rate, up to some order of $\log n$, as implied by the theoretical results below.

Despite the lack of a closed-form solution due to the
semi-positivity constraint and possibly non-differentiable penalties,  
we develop the empirical $L^2$ rate of convergence for covariance estimators in the tensor product Sobolev-Hilbert spaces.
This result is broad since it allows for a variety of spectral regularizations, including the trace-norm regularization and others, and incorporates both fixed and random designs.
Generally, the rate is comparable to
the optimal rate of standard two-dimensional nonparametric smoothers. If $X$ is additionally
periodic, we can improve our results significantly
  such that the optimal one-dimensional nonparametric rate, up to some order of $\log n$, is attained. 
For periodic functional spaces, when the data
are sparse, i.e., $m < \infty$, the rate of convergence is comparable to the
minimax rate obtained by \citet{Cai-Yuan10} and the $L^2$ rate achieved by
\citet{Paul-Peng09}. 
Different from these two pioneer works, our objective function is not necessarily
  differentiable, which thus requires separate theoretical treatments.
Our theoretical results are established in terms of empirical processes techniques. 
The success of the relevant proofs depends on the upper bound of the entropy for tensor
product Sobolev-Hilbert spaces, which is the first appearance in the FDA
literature to our best knowledge.

The rest of the paper is organized as follows.
The proposed methodology for covariance function estimation is presented in Section~\ref{sec:spec}. Computational issues and theoretical results are given in Sections~\ref{sec:computation} and~\ref{sec:asym} respectively. The empirical performance of the proposed approach is evaluated by a simulation study in Section~\ref{sec:sim} and a real data application in Section~\ref{sec:data}.
Additional materials, including technical details and further algorithmic descriptions, are provided in a separate supplemental document.

\section{Methodology}\label{sec:spec} 

\subsection{Spectral decomposition on RKHS}\label{subsec:RKHS}

We first introduce spectral decomposition on RKHS and then define a variety of spectral regularizations which we will use to obtain a class of covariance function estimators.

For a bivariate function $C(\cdot, \cdot)$ on $\mathcal{T}\times \mathcal{T}$, define its transpose, denoted by $C^\top$, as $C^\top(s,t)=C(t,s)$ for any $s,t\in\mathcal{T}$.
Due to the symmetry of covariance functions, we focus on the space $\SK=\{C\in\HKK: C = C^\top\}$.
For any $C\in\SK$,
define its self-adjoint operator $\mathcal{C}_C: \HK\rightarrow\HK$ by
\begin{equation}
  (\mathcal{C}_C f) (s) = \langle C(s,\cdot), f(\cdot) \rangle_{\HK}, \quad\quad \text{for any $f \in \HK$ and $s\in\mathcal{T}$.} \label{eqn:RKHSOp}
\end{equation}
Note that $\|C\|_{\HKK}<\infty$ since $C \in \SK$ and that the Hilbert-Schmidt norm of $\mathcal{C}_C$ coincides with $\|C\|_{\HKK}$. Therefore,
$\mathcal{C}_C$ is a Hilbert-Schmidt operator and hence
admits a spectral decomposition. In Section~\ref{subsec:spec}, we will define a penalty function based on this spectral decomposition.

In the FDA literature, the spectral analysis is often based on the Hilbert-Schmidt integral operator
$\mathcal{L}_C: L^2(\mathcal{T}) \rightarrow L^2(\mathcal{T})$ defined by
\begin{equation}
(\mathcal{L}_C f) (s) = \langle C(s,\cdot), f(\cdot)\rangle_{L^2(\mathcal{T})} = 
\int_\mathcal{T} C(s,t) f(t)\, dt, \quad \text{for any $f \in L^2(\mathcal{T})$ and $s\in\mathcal{T}$.} \label{eqn:L2Op}
\end{equation}

There are two reasons why we adopt $\mathcal{C}_C$ instead of $\mathcal{L}_C$. 
First, $\mathcal{C}_C$ is more aligned with the RKHS modeling of $X$, especially when the inner product of $\HK$ is chosen to mimic the physical reality.
Many examples can be found in the work on $L$-splines, e.g., Chapter~4.5 in \cite{Gu13} and Chapter~21 of \cite{Ramsay-Silverman05}.
Second, using $\mathcal{C}_C$ enables a finite dimensional representation
of our proposed covariance estimators as in Theorem \ref{thm:represent} below, and thus simplifies its practical computation.

\subsection{Spectrally regularized covariance estimator} \label{subsec:spec}

For any $C\in\SK$, let $\tau_1(C), \tau_2(C), \dots$ be the eigenvalues corresponding to the spectral decomposition of $\mathcal{C}_C$ such that $|\tau_1(C)|\ge|\tau_2(C)|\ge\cdots$.
We propose the following covariance estimator:
\begin{equation}
  \qquad \hat{C} = \argmin_{C\in\SpK} \left\{ \ell(C) + \lambda \Psi(C) \right\},
  \label{eqn:ggoal}
  \end{equation}
where $\SpK=\{ C\in\SK: \langle\mathcal{C}_C f, f\rangle_{\HK} \ge 0,\ \text{for all $f\in\HK$}\}$,
$\ell$ is a convex and smooth loss function, $\lambda>0$ is a tuning parameter, and $\Psi(C) = \sum_{k\ge1} \psi(|\tau_k(C)|)$ with 
$\psi$ being a non-decreasing penalty function satisfying $\psi(0)=0$ \citep{Abernethy-Bach-Evgeniou09}.
We assume that $\ell$ depends on $C$ through $\{C(T_{ij}, T_{ik}):i=1,\ldots,n; j, k =1, \ldots, m\}$.
The choice of $\psi$, and thus $\Psi$, is broad. In below we list a few interesting forms and briefly discuss their effects on the corresponding estimator.

\begin{ex}[Rank regularization] \label{ex:rank}
If $\psi(\tau) = I(\tau\neq 0)$ where $I(\cdot)$
    is the indicator function, $\Psi(C)$ is the rank of the operator $\mathcal{C}_C$. This penalty obviously encourages a low-rank solution.
    However, the minimization (\ref{eqn:ggoal}) is now difficult owing to its non-convexity, and 
over-fitting may occur since no regularizations are imposed on non-zero eigenvalues.
\end{ex}
\begin{ex}[Hilbert-Schmidt-norm regularization] \label{ex:HS} 
If $\psi(\tau)=\tau^2$, $\Psi(C)$ becomes the squared Hilbert-Schmidt norm of the operator $\mathcal{C}_C$, which equals $\|C\|_{\HKK}^2$.
Similar to the $\ell_2$-norm regularization for vectors, the Hilbert-Schmidt-norm regularization ensures the convexity of the objective function in (\ref{eqn:ggoal}), but does not encourage sparsity in eigenvalues, so the resulted covariance estimator is usually of high rank. \cite{Cai-Yuan10} used this regularization as in (\ref{eqn:CY}), but did not impose the constraint $C\in\SpK$, so a positive semi-definite covariance estimator was not guaranteed.
\end{ex}

\begin{ex}[Trace-norm regularization] \label{ex:trace}
If $\psi(\tau) = \tau$, 
    $\Psi(C)$ is the trace norm of $\mathcal{C}_C$, 
    which is a convex relaxation of its rank.
    Similar to the celebrated $\ell_1$-regularization for vectors
    and the trace-norm regularization for matrices
    the trace-norm penalty $\Psi$ for operators
    not only
    promotes the sparsity of eigenvalues and hence
    low-rank solutions, but also
   regularizes non-zero eigenvalues.
    The minimization (\ref{eqn:ggoal}) now becomes a convex optimization which allows leveraging
    many recent developments in non-smooth convex optimizations
    \citep[e.g.,][]{Beck-Teboulle09}
    to achieve feasible computations. See Section \ref{sec:computation}
    for more
    details.
\end{ex}

Obviously the penalty $\Psi(C)$ generalizes the regularization as in (\ref{eqn:CY}). Regardless of the form of $\Psi(C)$, the covariance estimator obtained by (\ref{eqn:ggoal}) is always positive semi-definite since the solution to the minimization (\ref{eqn:ggoal}) is searched only within $\SpK$.

\subsection{Representer theorem}\label{subsec:representer}

Since commonly used $\HK$, including $\mathcal{W}^r$, are infinite dimensional, solving (\ref{eqn:ggoal}) is typically an infinite dimensional optimization problem. 
Therefore, $\hat{C}$ is of little
practical value if a finite dimensional representation, based on data, is unavailable.  
To address this,
we provide a representer theorem which holds for the entire class
of estimators defined in (\ref{eqn:ggoal}).

Write $N=nm$ and $(\tilde{T}_1,\dots,\tilde{T}_N) =
(T_{11},\dots,T_{1m},T_{21}, \dots, T_{2m}, \dots, T_{n1},\dots, T_{nm})$. 

\begin{theorem}[Representer theorem]\label{thm:represent}
If the solution set of (\ref{eqn:ggoal}) is not empty, then there always exists a solution
lying in the space
  $\mathcal{K}\otimes\mathcal{K}=\mathrm{span}\{  K(\cdot, \tilde{T}_i)\otimes K(\cdot,
    \tilde{T}_{j}): i,j=1,\dots,N \}$,
where $\mathcal{K} = \mathrm{span}\{ K(\cdot, \tilde{T}_i): i=1,\dots,N \}$.
Moreover, the solution takes the form:
\begin{equation}
C(s,t) = z(s)^\top {A}\, z(t),\label{eqn:represent2}
\end{equation}
where $A$ is a $N\times N$ symmetric matrix and $z(\cdot) = (K(\cdot,\tilde{T}_1),\dots, K(\cdot, \tilde{T}_N))^\top$.
\end{theorem}
\noindent

Classical representer theorems \citep[e.g.,][]{Wahba90}, as adopted in
\citet{Cai-Yuan10}, do not
cover the scenario addressed by Theorem \ref{thm:represent} due to
the semi-positivity constraint and
a wide choice of regularizations, e.g., the trace-norm regularization.
To show this theorem, we significantly utilize the fact that the spectral analysis is based on
the RKHS geometry.
This is the main reason for using the operator $\mathcal{C}_C$
in (\ref{eqn:RKHSOp}) instead of
$\mathcal{L}_C$ in (\ref{eqn:L2Op}).
We remark that the conclusion of Theorem \ref{thm:represent} also holds when the semi-positivity is not imposed, i.e., $\mathcal{S}^{+}(K)$ is replaced by
$\mathcal{S}(K)$ in (\ref{eqn:ggoal}).
In Section \ref{sec:sim}, this fact will be used to compute unconstrained estimators 
for comparison.

At a first glance, a significant number of scalar parameters ($(N+1)N/2$) is involved in (\ref{eqn:represent2}). However, if a low-rank inducing penalty, such as the trace-norm regularization, is used,
the resulted estimator is often of low rank, which will benefit computation and storage in its estimation, and subsequent uses.
In Section~\ref{sec:computation}, we particularly focus on developing an efficient computational tool when the trace-norm regularization is imposed in (\ref{eqn:ggoal}).

\subsection{Parametrization}\label{subsec:para}
By Theorem \ref{thm:represent}, we are able to parametrize the solution to (\ref{eqn:ggoal}) in terms of a finite dimensional representation since 
it suffices to merely focus on covariance functions of the form $C(\cdot, \cdot) = \sum^N_{i=1}\sum^N_{j=1} {A}_{ij} K(\cdot, \tilde{T}_i)
\otimes K(\cdot, \tilde{T}_j)$.
The eigenvalues of the operator $\mathcal{C}_C$, $\{\tau_j(C): j \ge 1\}$,
are the eigenvalues of the matrix
$B=M^\top{A} M$,
where
$M$ is any $N\times \ra$ matrix such that $MM^\top=\tK
=[K(\tilde{T}_i,
\tilde{T}_j)]_{1\le i,j\le N}$ with $\ra=\mathrm{rank}(\tK)$.
The matrix $M$ provides a representation based on an orthonormal basis (of $\mathcal{K}$) 
$\{v_1,\dots, v_\ra\}$:
\[
  K(\cdot, \tilde{T}_i) \otimes K(\cdot, \tilde{T}_j) = \sum^\ra_{k=1}\sum^\ra_{l=1} M_{ik}M_{jl}
  v_k \otimes v_l, \quad\quad 1\le i,j\le N.
\]
Therefore $C=\sum^\ra_{k=1}\sum^\ra_{l=1} B_{kl} v_k \otimes v_l$.
As $C(s,t)=\langle K(\cdot,s), \mathcal{C}_C K(\cdot, t)\rangle_{\HK}$,
we have
  $[C(\tilde{T}_i,\tilde{T}_j)]_{1\le i, j\le N} = MBM^\top$.
Moreover, write
$M=(M_1^\top, \dots, M_n^\top)^\top$,
where $\{M_i: i=1, \ldots, n\}$ are $m\times \ra$ matrices.
Then the loss function depends on $C$ through $[C(T_{ij}, T_{ik})]_{1\le j, k\le
  m}=M_i B M_i^\top$ for $i=1,\dots,n$.
Compared with $A$, the new parametrization $B$ is unique even 
when $\{T_{ij}: i=1, \ldots, n; j=1, \ldots, m\}$ are not all unique.
Now (\ref{eqn:ggoal}) can be rewritten as
\begin{align}
\argmin_{B\in\mathcal{S}_\ra^{+}} \left\{\tilde{\ell}(B) + \lambda \tilde{\Psi}(B)\right\}, \label{eqn:goalB}
\end{align}
where
$\mathcal{S}_\ra^+$ is the set of all $\ra\times \ra$ positive semi-definite matrices,
$\tilde{\ell}(B) =
\ell(\sum^\ra_{k=1}\sum^\ra_{l=1} B_{kl} v_k \otimes v_l)$,
and $\tilde{\Psi}(B)=\sum_{k =1}^q\psi(|\xi_k(B)|)$ with $\xi_1(B),\dots, \xi_\ra(B)$
being the eigenvalues of the matrix $B$ such that $|\xi_1(B)|\ge\dots\ge|\xi_\ra(B)|$.
Conversely, with the new parametrization $B$, we can represent $C(s,t)=z(s)^\top (M^{+})^\top B M^{+} z(t)$ where
$M^+$ is the Moore-Penrose pseudoinverse of $M$.
Consequently, solving (\ref{eqn:ggoal}) is equivalent to solving (\ref{eqn:goalB}),
a finite-dimensional optimization.

\subsection[for the toc]{Closed-form expression of $L^2$ eigen-decomposition}\label{subsec:eigfun}
By Mercer's theorem, we can represent an arbitrary covariance function $C\in\SpK$ in terms of the typical spectral
decomposition via the $L^2$ inner product, i.e., 
$C(s,t) = \sum_{k\ge1} \zeta_k \phi_k(s)\phi_k(t)$,
where $\{\phi_k:k\ge 1\}$ are the $L^2$ eigenfunctions and
$\{\zeta_k: k\ge 1\}$ are the corresponding $L^2$ eigenvalues.
This eigen-decomposition is a key component of FPCA and other advanced FDA methods. In the literature \citep[e.g.,][]{Rice-Silverman91}, approximate computations are commonly involved where the eigen-decomposition is obtained based on the discretized covariance function estimator.  In contrast, due to Theorem~\ref{thm:represent}, our covariance estimator posseses a closed-form expression of this eigen-decomposition
so that such computational complication can be avoided.

Following the notations in Sections~\ref{subsec:representer} and \ref{subsec:para}, let $Q = [\int_\mathcal{T} K(s, \tilde{T}_i)K(s,\tilde{T}_j )\, ds]_{1\le i, j\le N}=M R M^\top$ where
$R = [\int_\mathcal{T} v_k(s)v_l(s)\, ds]_{1\le k,l \le \ra}$.
Note that $R = M^+ Q (M^+)^\top$.
Similar to Lemma~3 of \citet{Cai-Yuan10}, once $\hat{B}$ is obtained from (\ref{eqn:goalB}), 
the $L^2$ eigenfunctions of
$\hat{C}=\sum^\ra_{k=1}\sum^\ra_{l=1} \hat{B}_{kl} v_k \otimes v_l$
can be expressed as
  $\hat{\phi}_k(\cdot) = U^\top_k z(\cdot)$, for $k=1,\dots, n$,
where $U_k$ is the $k$-th column of $U=(M^+)^\top R^{-1/2} V$ and $V$ is the eigenvectors of
$R^{1/2} \hat{B} R^{1/2}$. 
The $L^2$ eigenvalues of $\hat{C}$ coincide with those of $R^{1/2} \hat{B} R^{1/2}$, and the number of nonzero eigenvalues is the same as the rank of $\hat{B}$.

\section{Computational issues for trace-norm regularization}\label{sec:computation}

To achieve a desirable low-rank covariance estimator, we develop an algorithm when the trace-norm regularization is used.

\subsection{Algorithm}\label{subsec:alg}
With the trace-norm regularization in (\ref{eqn:ggoal}), it is equivalent to solving the convex optimization
\begin{equation}
\argmin_{B\in\mathcal{S}_\ra^+} \left\{\tilde{\ell}(B) + \lambda \|B\|_*\right\}, \label{eqn:goalnuclear}
\end{equation}
where $\|\cdot\|_*$ represents the typical trace norm for matrices.
We can also rewrite (\ref{eqn:goalnuclear}) as
\begin{equation}
\argmin_{B\in\mathcal{S}_\ra} \left\{\tell(B) + \lambda h(B)\right\},
\quad\mbox{where}\quad
h(B) =
\begin{cases} \|B\|_*, & B\in\mathcal{S}_\ra^+ \\ \infty, &
  B\not\in\mathcal{S}_\ra^+ \end{cases}.
\label{eqn:goalnuclear2}
\end{equation}
Here $\mathcal{S}_\ra$ represents the set of all $\ra\times \ra$ matrices. 

The objective function in (\ref{eqn:goalnuclear2}) is the sum of a smooth function
and a non-smooth function.
A popular approach to such optimizations is the accelerated proximal gradient (APG) method \citep{Beck-Teboulle09}.
To apply this method, define an operator $\svec : \mathcal{S}_\ra \rightarrow \mathbb{R}^{\ra(\ra+1)/2}$ by
\[
  \svec(B) = [B_{11}, \sqrt{2}B_{21}, \dots, \sqrt{2}B_{\ra 1},
  B_{22}, \sqrt{2}B_{32}, \dots, \sqrt{2}B_{\ra 2},\dots,
  B_{\ra\ra}]^\top,
\]
for any $B=[B_{ij}]_{1\le i,j\le \ra}\in\mathcal{S}_\ra$.
This operator provides an isometry between $\mathcal{S}_\ra$ and $\mathbb{R}^{\ra(\ra+1)/2}$.
Denote its inverse by $\svec^{-1}$.
We write $\check{\ell}(b) = \tilde{\ell}(\mathrm{svec}^{-1}(b))$ for
any $b\in\mathbb{R}^{\ra(\ra+1)/2}$.
The APG algorithm of our case involves the proximal operator
$\mathrm{prox}_\nu: \mathbb{R}^{\ra(\ra+1)/2} \rightarrow \mathbb{R}^{\ra(\ra+1)/2}$ defined by
\begin{align*}
  \mathrm{prox}_{\nu} (b)
  &=\argmin_{d\in\mathbb{R}^{\ra(\ra+1)/2}} \left\{\frac{1}{2} \|d - b\|_E^2 + \nu h(\svec^{-1}(d))\right\}\nonumber\\
  &=\svec\left[\argmin_{D\in\mathcal{S}^+_\ra}\left\{ \frac{1}{2} \|D - B\|_F^2 + \nu\|D\|_*\right\}\right],
  \label{eqn:prox}
\end{align*}
for any $b\in\mathbb{R}^{\ra(\ra+1)/2}$ and $\nu>0$. Here $\|\cdot\|_E$ and $\|\cdot\|_F$
represent the Euclidean norm and the Frobenius norm respectively.
The following proposition states the closed-form solution of this proximal operator.

\begin{proposition}\label{prop:prox}
  For any $\nu>0$ and $b\in\mathbb{R}^{\ra(\ra+1)/2}$ with
  eigen-decomposition
  $\svec^{-1}(b)=P\mathrm{diag}(\tilde{b})P^\top$, 
  \[\mathrm{prox}_\nu (b) = \svec(P\mathrm{diag}(\tilde{c}) P^\top),\]
where
$\tilde{c}=( g_\nu(\tilde{b}_{1}),\dots,g_\nu(\tilde{b}_{\ra}))^\top$ and
$\tilde{b}=(\tilde{b}_{1},\dots,\tilde{b}_{\ra})^\top$. Here $g_\nu(x) = (x-\nu)_{+}$ for any $x\in\mathbb{R}$.
\end{proposition}
Due to this closed-form solution,
we can avoid the application of an inner numerical optimization within every
iteration of the APG algorithm.
The proof uses the same technique as in the proof of Lemma~1 in
\citet{Mazumder-Hastie-Tibshirani10}, and is thus omitted.

The standard APG method requires the knowledge of the Lipschitz
constant of $\nabla\check\ell$,
which directly relates to the step size in each iteration of
the algorithm.
For many choices of the loss function $\ell$, the corresponding Lipschitz
constant of $\nabla\check{\ell}$ is difficult to obtain.
Moreover, even when the Lipschitz constant is known (e.g., the choice of $\ell$ described
in Section \ref{sec:ell}),
the algorithm usually suffers from conservative
step sizes \citep{Becker-Candes-Grant11}.
Hence the APG with backtracking steps is usually preferred.
Following the suggestions of \cite{Becker-Candes-Grant11}, we adopt a modified version of the APG
method. See details in
Algorithm \ref{alg:trace_back}.
Modifying Steps 7--8 will result in other variants of
proximal gradient methods. See Section 5.2 of \citet{Becker-Candes-Grant11}
for more discussions. For convergence properties of the APG method, we refer interested readers to \cite{Beck-Teboulle09}.

  \linsps
\begin{algorithm}[h]
  \SetKwInOut{Input}{Input}
   \caption{The APG algorithm with backtracking for trace-norm-regularized covariance estimation }\label{alg:trace_back}
   \Input{$B_0\in\mathcal{S}_\ra^+$, $\hat{L}>0$, $\eta> 1$, $\alpha<1$}
   \BlankLine
    $b_0\leftarrow\mathrm{svec}(B_0)$, $\bar{b}_0 \leftarrow b_0$, $\theta_{-1}\leftarrow +\infty$, $L_{-1} \leftarrow \hat{L}$\\
    \For{$k=0,1,2,\dots$}{
     $L_k \leftarrow \alpha L_{k-1}$\\
     \Repeat{convergence}{
     $\theta_{k}\leftarrow 2/[1+ \{1+ 4L_k/ (L_{k-1}\theta_{k-1}^2)\}^{1/2}]$\\
     $e_k \leftarrow (1-\theta_k) b_k + \theta_k \bar{b}_k$\\
     ${b}_{k+1} \leftarrow \mathrm{prox}_{\lambda/L_k}(e_k - \nabla \check\ell(e_k) / L_k)$\\
     $\bar{b}_{k+1}\leftarrow \{b_{k+1} - (1-\theta_k)b_{k}\}/\theta_k$\\
     $\hat{L}\leftarrow 2|(e_k-b_{k+1})^\top \{\nabla\check\ell (b_{k+1})-\nabla\check\ell (e_{k})\}|/\|b_{k+1}-e_k\|_E^2$\\
     \If{$L_k\ge \hat{L}$}{break}
   $L_k \leftarrow \max\{ \eta L_k, \hat{L} \}$
 }
    }
\end{algorithm}
  \linsp

A similar algorithm can be obtained for (\ref{eqn:ggoal}) coupled
with the Hilbert-Schmidt-norm regularization, which 
will be implemented in Section~\ref{sec:sim} for a direct comparison.
Details are given in Section S1 of the supplemental document.

\subsection[for the toc]{A choice of $\ell$}\label{sec:ell}
Hereafter, we adopt the following {quadratic loss function}
\begin{equation}
  \ell(C) = \frac{1}{nm(m-1)} \sum^n_{i=1}\sum_{1\le j\neq k \le m}
  \left\{Z_{ijk} - C(T_{ij}, T_{ik})\right\}^2,  \label{eqn:lchoice}
\end{equation}
where
$Z_{ijk}=\{Y_{ij}-\hat{\mu}(T_{ij})\}\{Y_{ik}-\hat{\mu}(T_{ik})\}$ and
$\hat{\mu}$ is an estimator of the mean function $\mu_0$.
Since $[C(T_{ij},T_{ik})]_{1\le j,k\le m} = M_i B M_i^\top$ as shown in Section~\ref{subsec:para}, $\ell(C)$ becomes
\begin{align*}
  \tell(B) &= \frac{1}{nm(m-1)} \sum^n_{i=1}\| \rho(Z_i - M_i B M_i^\top)\|^2_F
  = \frac{1}{2} \mathrm{vec} (B)^\top \nabla^2 \tell(B) \mathrm{vec}(B) -
  \left(\sum^n_{i=1} M_i^\top Z_i M_i\right)^\top\mathrm{vec}(B),
\end{align*}
up to an additive constant independent of $B$. 
Here $Z_i=[Z_{ijk}]_{1\le j,k\le m}$, $\rho$ is an operator setting the
diagonal entries of its input to zero,
and
\[
  \nabla^2 \tell(B) = \frac{2}{nm(m-1)} \sum^n_{i=1} (M_i^\top \otimes M_i^\top)
  \mathrm{diag}\{\mathrm{vec}(\tilde{I})\} (M_i \otimes M_i),
\]
with $\tilde{I}\in \mathbb{R}^{q\times q}$ consisting of elements $\tilde{I}_{ij} = I(i\neq j)$. 
With straightforward derivations, one can obtain the closed-form expressions 
of $\check{\ell}$ and $\nabla\check\ell$ as required in Algorithm \ref{alg:trace_back}, which we omit here.

\section{Asymptotic properties}\label{sec:asym}
In this section, we develop the empirical $L^2$ rate of convergence for a variety of spectrally regularized covariance estimators in the tensor product Sobolev-Hilbert spaces. This result is broad since it incorporates both fixed and random designs, and also allows for a variety of spectral regularizations, including the trace-norm, Hilbert-Schmidt regularizations and others.
\subsection{Assumptions} \label{subsec:assump}
Without loss of generality, we take $\mathcal{T} = [0, 1]$. Below we establish 
the asymptotic properties for the $r$-th order Sobolev-Hilbert space on $[0, 1]$ where $r \ge 2$, i.e., 
\[
\HK = \{g: g^{(v)}, v=0, \ldots, r-1, \,\text{are absolutely continuous};\, g^{(r)} \in L^2([0, 1])\}.
\]
The space $\HK$ is equipped with squared norm $\|g\|^2= \sum^{r}_{v=0} \int^1_0 \{g^{(v)}(t)\}^2\,dt $.
The asymptotic results also hold for its equivalent norms, e.g., $\|g\|^2= \int^1_0 \{g(t)\}^2\, dt + \int^1_0 \{g^{(r)}(t)\}^2\, dt$, $\|g\|^2=( [\int^1_0 \{g(t)\}^2\, dt]^{1/2} + [\int^1_0 \{g^{(r)}(t)\}^2\, dt]^{1/2} )^2$, and $\|g\|^2= \sum^{r-1}_{v=0} \{ \int^1_0 g^{(v)}(t)\,dt\}^2 + \int^1_0 g^{(r)}(t)^2\, dt$.

We list the assumptions needed for the asymptotic properties as follows.

 \begin{assumption} \label{assum:C0} 
 $C_0 \neq 0$ and $C_0 \in\mathcal{F}
 \subseteq \HKK$ where $\mathcal{F}$ is the hypothesis space for estimation.
 \end{assumption}
  \begin{assumption} \label{assum:T} 
The time points $\{T_{ij}: i=1,\ldots, n; j=1,\ldots, m\}$ are either fixed or random, and are independent of $\{X_i: i=1, \ldots, n\}$. The errors $\{\varepsilon_{ij}: i=1,\dots, n; j=1,\dots m\}$ are independent of both $\{T_{ij}: i=1,\ldots, n; j=1,\ldots, m\}$ and $\{X_i: i=1, \ldots, n\}$.
 \end{assumption}
 \begin{assumption} \label{assum:X} 
 For each $t\in [0, 1]$, $X(t)$ is sub-Gaussian with a parameter $b_X>0$ which does not depend on $t$, i.e., $\E(\exp\{\beta X(t)\})\le \exp\{b_X^2 \beta^2/2\}$ for all $\beta>0$ and $t\in [0, 1]$.
 \end{assumption} 
 \begin{assumption} \label{assum:varesp} 
   For each $i, j$, $\vareps_{ij}$ is sub-Gaussian with a parameter $b_\vareps$ independent of $i$ and $j$.
 \end{assumption} 

As shown in \citet{Cai-Yuan10},
$C_0\in\HKK$
under the assumptions that $X\in\mathcal{H}(K)$ almost surely and
$\E\|X\|^2_{\mathcal{H}(K)}<\infty$.
Assumption~\ref{assum:T} is standard in FDA modeling.
Assumptions~\ref{assum:X} and~\ref{assum:varesp} are sub-gaussian conditions
of the stochastic process and the measurement error.

\subsection{Rate of convergence}\label{subsec:rate}
We investigate the asymptotic property of a class of covariance estimators given by
\begin{equation}\label{eqn:ggoalnew}
\hat{C}_{\lambda} = \argmin_{C\in \mathcal{F}} \left\{ \ell(C) + \lambda \Psi(C) \right\}, 
\end{equation}
where $\Psi(C) =\sum_{k\ge1} |\tau_k(C)|^p$ for $1 \leq p \leq 2$
and the loss function $\ell$ is chosen as (\ref{eqn:lchoice}).
Apparently, the penalty term $\Psi$ generalizes both trace-norm ($p=1$) and Hilbert-Schmidt-norm ($p=2$) regularizations. Moreover, $\hat{C}_{\lambda}$ becomes the estimator by \citet{Cai-Yuan10} if $\mathcal{F} = \HKK$ and $p=2$.
For simplicity, we assume known $\mu_0=0$ so we let $\hat{\mu}=0$ and accordingly $Z_{ijk}=Y_{ij} Y_{ik}$. 

For arbitrary bivariate functions $g_1$ and $g_2$, define an empirical inner product and the corresponding empirical norm as follows:
\[
\ip{g_1, g_2}_n = \frac{1}{nm(m-1)}\sum_{i=1}^n \sum_{1 \leq j \neq k \leq m} g_1(T_{ij}, T_{ik})g_2(T_{ij}, T_{ik}) \quad\mbox{and}\quad \norm{g_1}_n^2 = \ip{g_1, g_1}_n.
\]
Recall that we say a random variable $S_n=\bigO_{p}(k_n)$ if
    \[
    \lim_{L\rightarrow\infty} \limsup_{n \rightarrow \infty}\, \pr(S_n\ge L k_n) =0.
    \]
  To accommodate the flexibility of the design $\mathbb{T}=\{T_{ij}: i=1, \ldots, n; j=1, \ldots, m\}\in\mathcal{T}^{nm}$, we denote $S_n=\bigO_{p}^T(k_n)$ if
  \[
    \lim_{L\rightarrow\infty} \limsup_{n \rightarrow \infty}
    \sup_{\mathbb{T}\in\mathcal{T}^{nm}}\, \pr(S_n\ge L k_n \mid \mathbb{T}) =0.
  \]
  
We first provide the empirical $L^2$ rate of convergence for $\hat{C}_{\lambda}$. 
\begin{theorem}\label{thm:2drate}
Under Assumptions~\ref{assum:C0}--\ref{assum:varesp},
if $\Psi(C_0) > 0$ and $\lambda^{-1} = \bigO_p \left\{ n^{r/(1+r)}\right\}$,
we have 
$\norm{\hat{C}_{\lambda} - C_0}_n =\bigO_p( \lambda^{1/2} )$.
Further, if
$\lambda^{-1} = \bigO_p^T \left\{ n^{r/(1+r)}\right\}$,
we have 
$\norm{\hat{C}_{\lambda} - C_0}_n =\bigO_p^T( \lambda^{1/2} )$.
\end{theorem}
In Theorem \ref{thm:2drate}, the asymptotic accuracy of $\hat{C}_\lambda$ is guaranteed for both fixed and random designs. In particular, both independent and dependent designs are also allowed if the design is random. Furthermore, Theorem \ref{thm:2drate} provides a uniform result over all designs under a stronger condition of $\lambda$. For instance, such $\bigO_p^T$-condition degenerates to the weaker $\bigO_p$-condition if the choice of $\lambda$ is nonrandom or independent of the design.

Theorem~\ref{thm:2drate} incorporates a variety of regularizations as long as
$1 \leq p \leq 2$, where the commonly used trace-norm ($p=1$) and
Hilbert-Schmidt-norm ($p=2$) penalties are both special cases. It shows that
the empirical $L^2$ rate of convergence of $\hat{C}_{\lambda}$ is comparable to
that of standard two-dimensional nonparametric smoothers. For example, the rate
of convergence is $n^{1/3}$ for the second order Sobolev-Hilbert space, i.e.,
$r=2$.
The conclusion in Theorem~\ref{thm:2drate} is generally true for all two-dimensional Sobolev spaces, but the rate is sub-optimal within the scope of tensor product Sobolev-Hilbert spaces. 
For periodic functions, however, we are able to significantly improve this rate by utilizing appropriate and specific entropy results for tensor product Sobolev-Hilbert spaces.

\begin{theorem}\label{thm:rate}
Suppose that $\mathcal{F} \subseteq \{ C \in \HKK: \text{$C$ is a periodic function}\}$. Under Assumptions~\ref{assum:C0}--\ref{assum:varesp},
if $\Psi(C_0) > 0$, and
$\lambda^{-1} = \bigO_p \{ {n^{2r/(1+2r)}}/{\log n}\}$,
we have 
$\norm{\hat{C}_{\lambda} - C_0}_n = \bigO_p( \lambda^{1/2} )$.
Further, if
$\lambda^{-1} = \bigO_p^T \{ {n^{2r/(1+2r)}}/{\log n}\}$,
we have 
$\norm{\hat{C}_{\lambda} - C_0}_n = \bigO_p^T( \lambda^{1/2} )$.
\end{theorem}

Similar to Theorem~\ref{thm:2drate}, Theorem~\ref{thm:rate} also allows for both fixed and random designs.
Theorem~\ref{thm:rate} demonstrates that $\hat{C}_{\lambda}$ can achieve the empirical $L^2$ rate of convergence for one-dimensional nonparametric estimation, up to some order of $\log n$, although the target function $C_0$ is two-dimensional.
For instance, if we let $r=2$, the rate of $\hat{C}_{\lambda}$ is $(\log n)^{-1/2} n^{2/5}$, which is much faster than the two-dimensional nonparametric rate $n^{1/3}$.
For sparse functional data, i.e., $m < \infty$, up to some order of $\log n$, the rate of $\hat{C}_{\lambda}$ is comparable to the minimax rate obtained by \citet{Cai-Yuan10} and the $L^2$ rate achieved by \citet{Paul-Peng09} for $r=4$. However, the rates in both theorems are sub-optimal for functional data that are not sparse \citep{ZhanW16}. 

The covariance estimator $\hat{C}_{\lambda}$ defined in (\ref{eqn:ggoalnew}) does not have a closed form due to the possible non-differentiability of the penalty term (e.g., when $p=1$), and the flexibility of $\mathcal{F}$. This explains the technical challenges and highlights the novelties of the proofs for Theorems~\ref{thm:2drate} and~\ref{thm:rate}.
In Theorem~\ref{thm:rate}, the particular structure of the tensor product RKHS accounts for the appealing rate of convergence of $\hat{C}_{\lambda}$. The upper bound of the entropy for tensor product Sobolev-Hilbert spaces, as given in Lemma~1 of the supplemental document, is a crucial component for the technical success. To our best knowledge, this paper is the first one in the FDA literature that achieves this result.

\section{Simulation experiments}\label{sec:sim}

Numerical experiments were conducted to illustrate the practical performance of
the proposed methodology.
We generated $\{X_i: i=1, \ldots, n\}$ where $n=200$ from a Gaussian process with $\mu_0(t) = 3\sin\{3\pi (t + 0.5)\} + 2 t^3$
and $C_0(s, t) = \sum_{k=1}^L (k+1)^{-2} \phi_k(s) \phi_k(t)$,
with $\phi_1(t) = 2^{1/2} \cos(2 \pi t)$, $\phi_2(t) = 2^{1/2} \sin(2 \pi t)$, $\phi_3(t) = 2^{1/2} \cos(4 \pi t)$, and $\phi_4(t) = 2^{1/2} \sin(4 \pi t)$.
We also sampled $\{T_{ij}: i=1,\dots, n; j=1,\dots m\}$ independently from the uniform distribution on $[0, 1]$ and $\{\varepsilon_{ij}: i=1,\dots, n; j=1,\dots m\}$ independently from $N(0, 0.01)$ to produce $Y_{ij} = X_i(T_{ij}) + \varepsilon_{ij}$. 
We studied six settings in total, where $L=2$ or $4$, and $m=5$, $10$ or $20$. 
In each setting, we simulated 300 datasets
where we compared various covariance function estimators.
Other than our proposed estimators, we also included popular alternatives, including the
covariance smoothing estimators by local polynomial regression \citep{Yao-Muller-Wang05} and bivariate P-splines \citep{Goldsmith-Bobb-Crainiceanu11} respectively.
In these common alternatives, a raw smoothed estimate is first computed. Then
a truncation step via FPCA is often applied to reconstruct a covariance function
that is both positive semi-definite and of low rank.
That means, the reconstructed covariance estimator, which we refer to as a two-step estimator below,
takes the form: $\sum_{k=1}^J \hat{\zeta}_k\hat{\phi}(s)\hat{\phi}_k(t)$ where $\hat{\zeta}_k$'s and $\hat{\phi}_k$'s are the largest $J$ positive estimated eigenvalues and corresponding estimated eigenfunctions based on the raw smoothed estimate.

Altogether we compared the following ten methods, {of which the first five are based on our proposed framework while the rest are popular alternatives}:
1) $\hat{C}^{+}_{\mathsf{trace}}$: {obtained from} (\ref{eqn:ggoal}) with trace-norm regularization;
2) $\hat{C}_{\mathsf{trace}}$: {obtained from} (\ref{eqn:ggoal}) with trace-norm regularization but {without}
the semi-positivity constraint, i.e., $\mathcal{S}^+(K)$ replaced by $\mathcal{S}(K)$;
3) $\hat{C}^{+}_{\mathsf{HS}}$: {obtained from} (\ref{eqn:ggoal}) with Hilbert-Schmidt norm regularization;
4) $\hat{C}_{\mathsf{HS}}$: {obtained from} (\ref{eqn:ggoal}) with
    Hilbert-Schmidt norm regularization but {without}
    the semi-positivity constraint, which is exactly (\ref{eqn:CY});
5) $\hat{C}_{\mathsf{CY}}$: the estimator proposed by \cite{Cai-Yuan10} as defined by (\ref{eqn:CY}) and
    implemented in their R package\footnote{Downloaded from \url{http://stat.wharton.upenn.edu/~tcai/paper/html/Covariance-Function.html}};
6) $\hat{C}_{\mathsf{PACE}}$: the raw smoothed covariance estimator using local polynomial regression
\citep{Yao-Muller-Wang05} implemented in the R package \textsf{fdapace};
7) $\hat{C}_{\mathsf{PACE,FVE}}^{+}$: the two-step estimator based on $\hat{C}_{\mathsf{PACE}}$ with the 
number of components $J$ selected by fraction of variation explained (\textsc{fve})\footnote{With package default threshold 0.9999}, implemented in \textsf{fdapace};
8) $\hat{C}_{\mathsf{PACE,AIC}}^{+}$: the two-step estimator based on $\hat{C}_{\mathsf{PACE}}$ with
$J$ selected by Akaike Information Criterion (\textsc{aic}), implemented in \textsf{fdapace};
9) $\hat{C}_{\mathsf{PACE,BIC}}^{+}$: the two-step estimator based on $\hat{C}_{\mathsf{PACE}}$ with
$J$ selected by Baysian Information Criterion (\textsc{bic}), implemented in \textsf{fdapace};
10) $\hat{C}_{\mathsf{SC}}^{+}$: the two-step estimator based on tensor product
bivariate P-spline smoothing \citep{Goldsmith-Bobb-Crainiceanu11}
with $J$ selected by \textsc{fve}\footnote{With package default threshold 0.99},
implemented in the R package \textsf{refund}. 
Note that \textsf{refund} does not output the raw smoothed estimate of $\hat{C}_{\mathsf{SC}}^{+}$ and therefore we do not compare such raw estimator.

{To obtain $\hat{C}^{+}_\mathsf{trace},\hat{C}_\mathsf{trace},\hat{C}^{+}_\mathsf{HS},\hat{C}_\mathsf{HS}$
and $\hat{C}_\mathsf{CY}$,}
smoothing spline was first applied to estimate $\mu$, where its smoothing parameter was selected by generalized cross-validation (\textsc{gcv}).
To obtain each of these five covariance estimators, we always used the loss function $\ell$ as in Section~\ref{sec:ell}, and $\HK$ was chosen as the second order Sobolev-Hilbert space on
$[0,1]$ with the squared norm
$\|g\|^2 = \sum^{1}_{v=0} \{ \int^1_0 g^{(v)}(t)dt\}^2 + \int^1_0 \{g^{(2)}(t)\}^2 dt$.
The tuning parameter $\lambda$ of the first four methods were chosen by
five-fold cross-validation.
The computations of $\hat{C}_{\mathsf{trace}}$, $\hat{C}^+_{\mathsf{HS}}$ and $\hat{C}_{\mathsf{HS}}$
were achieved by Algorithm \ref{alg:trace_back}
with different proximal operators (Step 7) due to the change of penalty and
the utility of positivity constraint.
For the remaining five methods, $\mu$ is estimated by the corresponding computational packages. {See their documents for further implementation details.}

Table \ref{tab:rkhscov} shows the average integrated squared errors (\textsc{aise})
and
average ranks of these covariance estimators over 300 simulated data sets.
First, we restrict our attention to the first five methods which can all be regarded as variations of (\ref{eqn:ggoal}).
Although $\hat{C}_{\mathsf{CY}}$ and $\hat{C}_{\mathsf{HS}}$ share the same definition (\ref{eqn:CY}), they differ in various implementation details
and hence
the practical performance.
However, their differences in \textsc{aise}
{are too small to}
affect the subsequent {comparisons in this section, so hereafter it suffices to include only $\hat{C}_{\mathsf{HS}}$, rather than both of them, to study rank reduction and the effect of the semi-positivity constraint. }

When we compare the two pairs, $\hat{C}^{+}_\mathsf{trace}$ versus $\hat{C}_\mathsf{trace}$, and $\hat{C}^{+}_\mathsf{HS}$ versus $\hat{C}_\mathsf{HS}$, obviously the covariance estimators with the positivity constraint always achieve smaller \textsc{aise} values than their counterparts.
This suggests that not only can imposing the semi-positivity constraint produce a valid estimator, but also improve estimation accuracy.
Notice in Table \ref{tab:rkhscov} that rank reduction can also be observed for
$\hat{C}^{+}_\mathsf{HS}$ because the semi-positivity constraint often results
in truncations of eigenvalues at zero.
When $\hat{C}^{+}_\mathsf{trace}$ is compared with $\hat{C}^{+}_\mathsf{HS}$, the former performs slightly worse in \textsc{aise}, but significantly better in rank reduction. The average ranks of $\hat{C}^{+}_\mathsf{trace}$ are in fact the closest to the true rank among the three rank-reduced estimators.
This highlights the benefits of trace-norm regularizations in computation, storage and subsequent uses as mentioned in Section~\ref{sec:intro}. {Next we only compare $\hat{C}^{+}_\mathsf{trace}$ with the five popular alternatives.} 
{The three two-step estimators based on local polynomial regression, $\hat{C}_{\mathsf{PACE,FVE}}^{+}$, $\hat{C}_{\mathsf{PACE,AIC}}^{+}$ and $\hat{C}_{\mathsf{PACE,BIC}}^{+}$, have similar average ranks, which explains why their \textsc{aise} values are almost identical. When compared with the raw smooth estimator $\hat{C}_{\mathsf{PACE}}$, they can all reduce rank and achieve smaller \textsc{aise} values, which illustrates the benefit of the reconstruction step of the classical two-step procedure. 
However, their average ranks and  \textsc{aise} values are both much higher than those of $\hat{C}_{\textsf{trace}}^{+}$ in all settings. Thus our proposed $\hat{C}_{\textsf{trace}}^{+}$ uniformly and significantly outperforms all these estimators with respect to both rank reduction and estimation accuracy.} 

{When compared with $\hat{C}_{\mathsf{SC}}^{+}$, $\hat{C}_{\textsf{trace}}^{+}$ achieves similar \textsc{aise} values and performs slightly but uniformly better in rank reduction in all six settings. Table \ref{tab:rkhscov} also shows that $\hat{C}_{\textsf{trace}}^{+}$ is numerically more stable than $\hat{C}_{\mathsf{SC}}^{+}$. Computational errors occurred in some simulation runs to obtain $\hat{C}_{\mathsf{SC}}^{+}$, but none appeared for $\hat{C}_{\textsf{trace}}^{+}$. Therefore, $\hat{C}_{\textsf{trace}}^{+}$ is a more desirable covariance estimator than $\hat{C}_{\mathsf{SC}}^{+}$.} 

{In summary, the overall performance of $\hat{C}_{\textsf{trace}}^{+}$ is the best among the ten covariance estimators regarding rank reduction, estimation accuracy, and computational stability. This motivates us to use $\hat{C}^{+}_\mathsf{trace}$ in the following real data application.}

\linsps
\begin{table}[ht]
\centering
{\small  \caption{\textsc{aise} ($\times 10^3$) values with standard errors ($\times 10^3$) in parentheses for the {ten} covariance estimators, 
and average ranks for those estimators with rank reduction. {For $\hat{C}^{+}_{\mathsf{SC}}$, whenever computational errors occurred in a setting, its statistics are computed only based on successful runs, with the proportion of success runs additionally shown in square brackets.}
  }
  \label{tab:rkhscov}
\begin{tabular}{rrr|rrrrr}
    \hline\hline
$m$ & $L$ &   & $\hat{C}^{+}_\mathsf{trace}$ & $\hat{C}_\mathsf{trace}$ &
    $\hat{C}^{+}_\mathsf{HS}$ & $\hat{C}_\mathsf{HS}$ & $\hat{C}_\mathsf{CY}$ \\\hline
5 & 2 & \textsc{aise} &  4.86 (0.168) &  7.64 (0.159) &  4.77 (0.159) &  8.22 (0.171) &  6.43 (0.146) \\ 
  && rank &2.77 & 6.14 & 14.8 & - & - \\
  & 4 & \textsc{aise} &  9.66 (0.262) & 14.98 (0.244) &  9.15 (0.237) & 15.54 (0.220) & 12.26 (0.293) \\ 
  && rank  & 3.90 & 7.28 & 15.3 & - & - \\
  \hline
10 & 2 & \textsc{aise} &  2.85 (0.0918) &  3.56 (0.0953) &  2.81 (0.0888) &  4.08 (0.0904) &  4.30 (0.0825) \\ 
 && rank  & 2.74 & 8.00 & 14.5 & - & - \\
& 4 & \textsc{aise} &  4.94 (0.107) &  7.97 (0.123) &  4.78 (0.0991) &  8.25 (0.125) &  7.25 (0.104) \\ 
&&rank  & 4.44 & 12.9 & 14.9 & - & - \\
  \hline
20 & 2& \textsc{aise} &  2.07 (0.0777) &  2.27 (0.0782) &  2.06 (0.0772) &  2.54 (0.0782) &  3.46 (0.0787) \\ 
 && rank  & 2.68 & 8.47 & 14.3 & - & - \\
 & 4 & \textsc{aise} &  3.27 (0.0809) &  4.69 (0.0846) &  3.24 (0.0808) &  4.69 (0.0853) &  6.04 (0.0764) \\ 
&& rank  & 4.52 & 22.0& 15.0 & - & - \\
   \hline\hline
$m$ & $L$ &   & $\hat{C}_{\mathsf{PACE}}$ & $\hat{C}_{\mathsf{PACE,FVE}}^{+}$ & $\hat{C}_{\mathsf{PACE,AIC}}^{+}$ & $\hat{C}_{\mathsf{PACE,BIC}}^{+}$ & $\hat{C}^{+}_{\mathsf{SC}}$\\
  \hline
5 & 2 & \textsc{aise}  &  8.99 (0.1940) &  8.29 (0.1976) &  8.29 (0.1976) &  8.29 (0.1976) &  5.20 (0.1715) [90.0\%] \\
  &   & rank & - & 6.57 & 5.58 & 5.57 & 4.15 [90.0\%] \\
 & 4 & \textsc{aise} & 11.54 (0.2016) & 10.71 (0.2030) & 10.71 (0.2030) & 10.71 (0.2030) &  7.98 (0.3392) [77.7\%] \\
  &   & rank & - & 7.16 & 6.16 & 6.16 & 5.21 [77.7\%] \\\hline
  10 & 2 & \textsc{aise} &  6.69 (0.1295) &  6.37 (0.1264) &  6.37 (0.1264) &  6.37 (0.1264) &  3.18 (0.0974) \\
    &   & rank & - & 5.92 & 4.95 & 4.93 & 4.14 \\
 & 4 & \textsc{aise} & 9.01 (0.1393) &  8.60 (0.1360) &  8.60 (0.1360) &  8.60 (0.1360) &  4.65 (0.1035) \\
    &   & rank & - & 6.46 & 5.50 & 5.49 & 5.35 \\\hline
  20 & 2 & \textsc{aise} &  5.81 (0.1159) &  5.56 (0.1118) &  5.56 (0.1118) &  5.56 (0.1118) &  2.23 (0.0804) \\
    &   & rank & - & 5.35 & 4.40 & 4.38 & 3.86 \\
 & 4 & \textsc{aise} & 7.88 (0.1205) &  7.58 (0.1160) &  7.58 (0.1160) &  7.58 (0.1160) &  3.14 (0.0812) \\
    &   & rank & - & 5.99 & 5.08 & 5.04 & 5.09 \\
   \hline\hline
\end{tabular}}

\end{table}

\linsp

\section{Real data application}\label{sec:data}

We apply the proposed method to a loop sensor dataset which 
contains vehicle counts recorded every five minutes at an on-ramp on the 101
North freeway in Los Angeles, U.S.A.. This on-ramp is located near Dodger Stadium, the home field of the Los Angeles Dodgers
baseball team,
so unusual traffic is expected after a Dodgers home game.
These measurements were collected by the Freeway Performance Measurement System
(PeMS)\footnote{\url{http://pems.dot.ca.gov}}
and can be obtained from the UCI Machine Learning Repository\footnote{\url{https://archive.ics.uci.edu/ml/datasets/Dodgers+Loop+Sensor}}.
We focus on the after-game traffic measurements of 78 games between April 2005 and October 2005 available in this dataset.
For each game, we have 31 measurements that cover the time interval from 30 minutes
before the
end of the game, to 120 minutes after the end of the game.
This time interval is presented as $[-30, 120]$, where zero
marks the end of a game.

\linsps
\begin{figure}[ht]
  \centerline{
    \includegraphics[height=7cm]{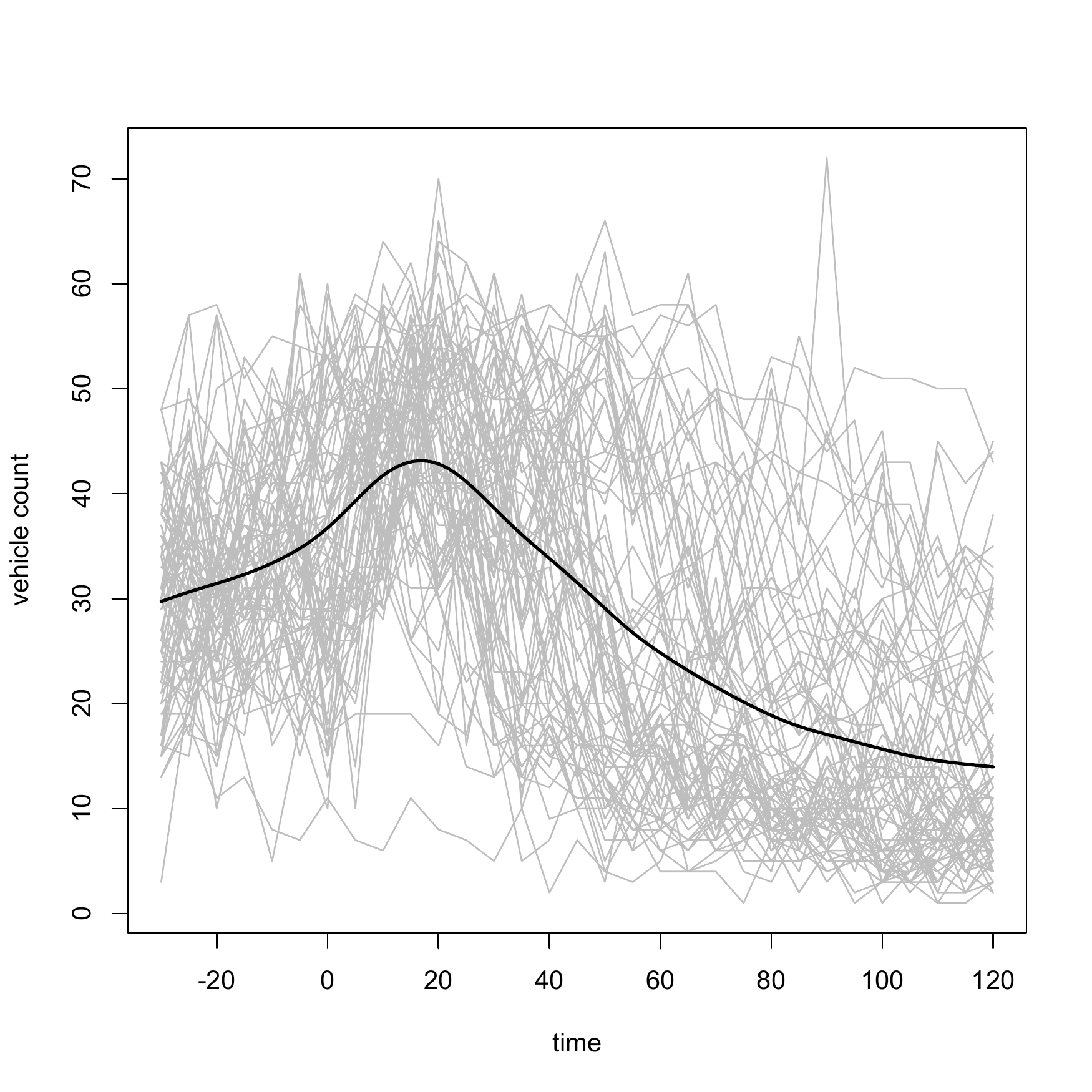}
  }
  \caption{Vehicle counts over a time interval from 30 minutes before the end
    of a game, to 120 minutes after the end of the game.
    The black line represents a  smoothing spline estimate of the mean function.}\label{fig:real:data}
\end{figure}
\linsp

The vehicle counts of the 78 games are displayed
in Figure \ref{fig:real:data}, where the mean function was estimated by smoothing splines with its tuning parameter determined by \textsc{gcv}.
The estimated mean curve demonstrates a traffic peak that 
emerges at around 20 minutes after the end of a game. This characteristic is consistent with the finding of \citet{Zhang-Wang15} and conforms to common sense.

\linsps
\begin{figure}[ht]
  \centerline{
    \includegraphics[height=6cm]{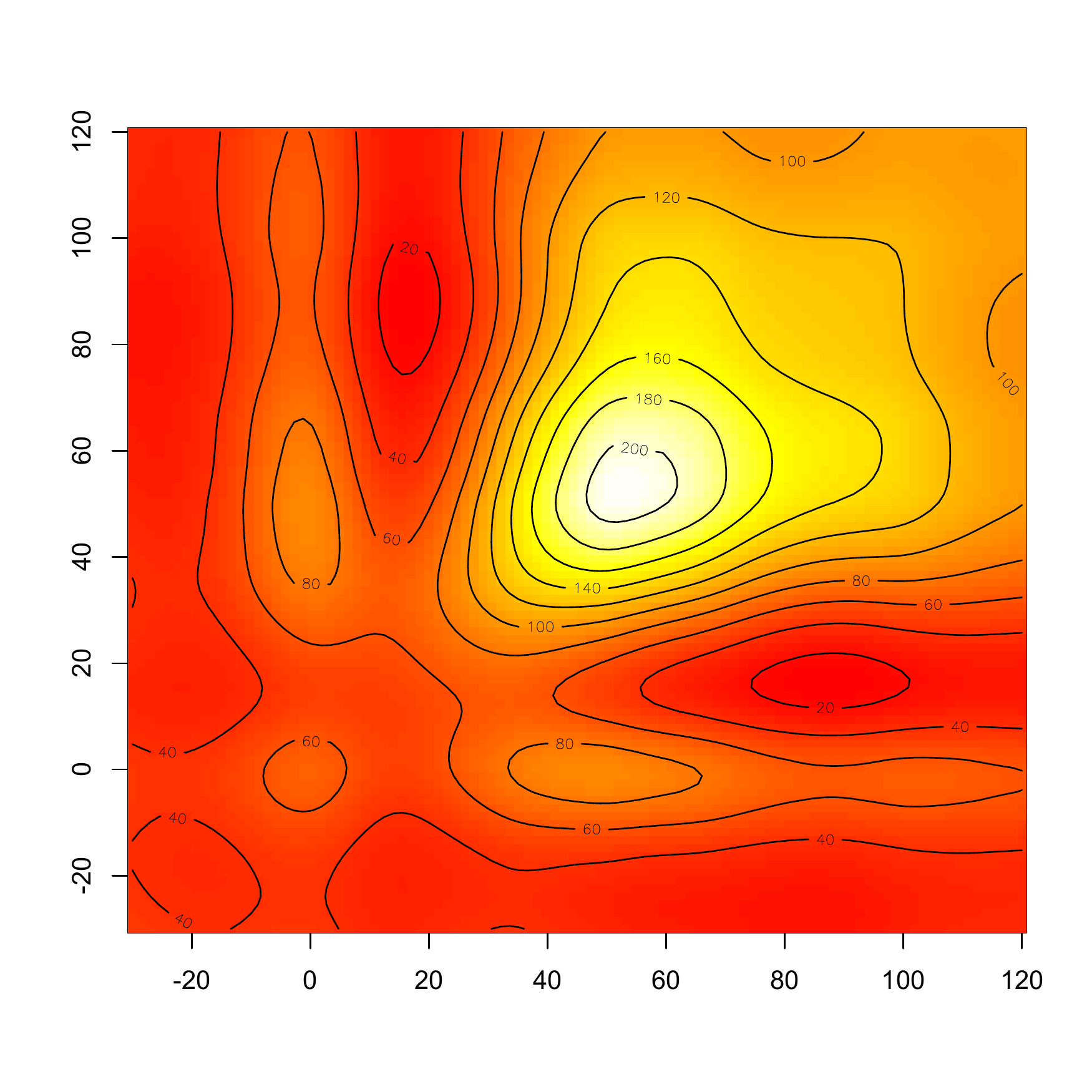}
    \includegraphics[height=6cm]{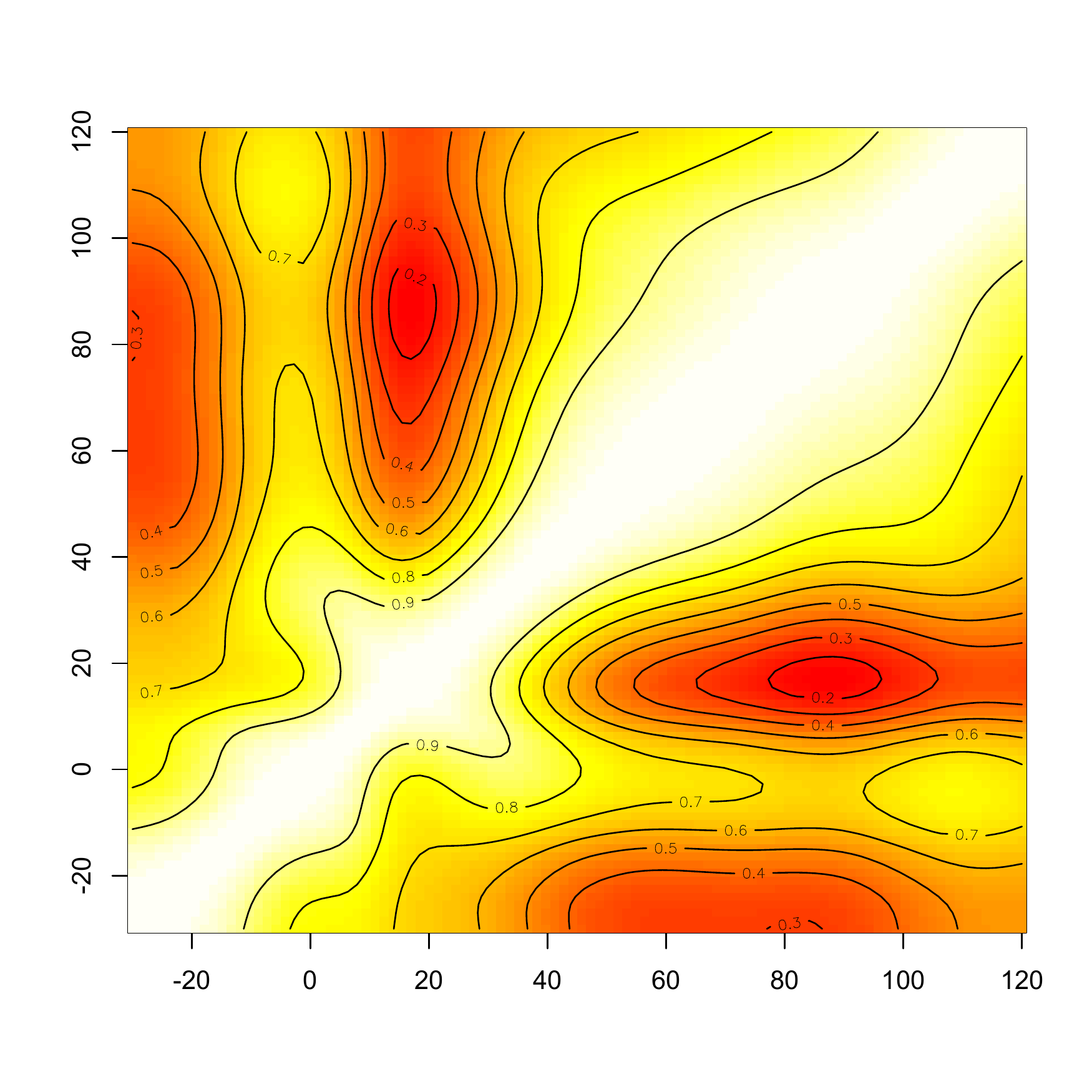}
  }
  \caption{Contour plots of the estimated covariance (left) and correlation (right) functions respectively for $\hat{C}^{+}_\mathsf{trace}$.}\label{fig:real:contour}
\end{figure}
\linsp

We provided the covariance estimator $\hat{C}^{+}_\mathsf{trace}$ of the vehicle counts, as described in Section \ref{sec:sim},
and constructed the corresponding correlation function estimate by the simple transformation: $\hat{C}(s,t)/\{\hat{C}(s,s) \hat{C}(t,t)\}^{1/2}$ for any covariance estimate $\hat{C}$ with $\hat{C}(s,s)>0$ for all $s$.
Note that positive semi-definiteness guarantees the validity of 
the correlation function estimate obtained by the above simple transformation.
Namely, it has value in-between -1 and 1.
However, this property could be violated for non-positive semi-definite
estimators such as $\hat{C}_\mathsf{CY}$.
The covariance and correlation estimate for $\hat{C}^{+}_\mathsf{trace}$ are depicted in Figure \ref{fig:real:contour}.
One intriguing feature with respect to the temporal dependency of the vehicle counts
is the high correlations of traffic between time $0$ and
time points after around time $30$.
When compared with adjacent time points such as $-20$ and $20$,
this feature is so distinctive that a ridge is formed at time $0$.

\begin{figure}[ht]
  \centerline{
    \includegraphics[height=6cm]{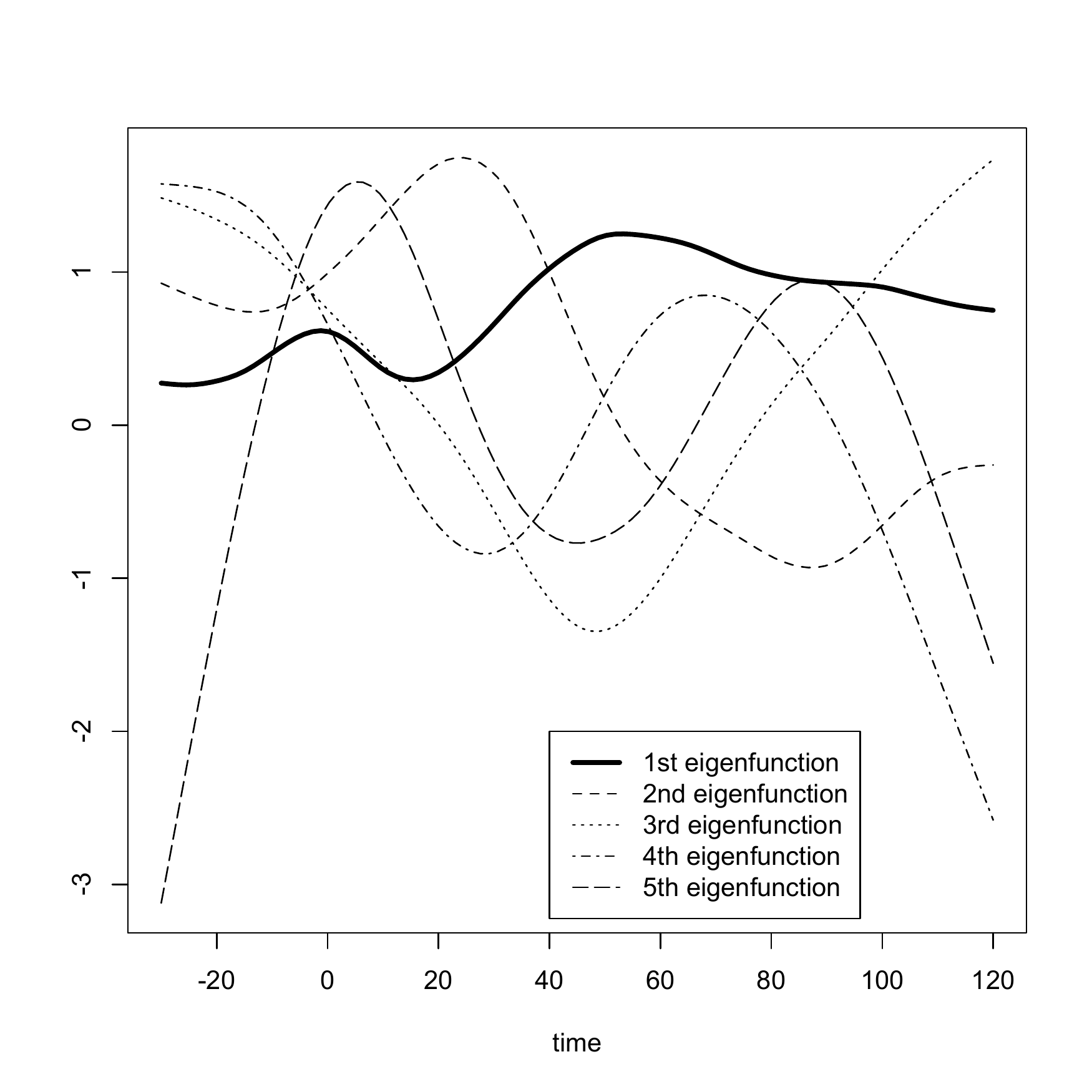}
    \includegraphics[height=6cm]{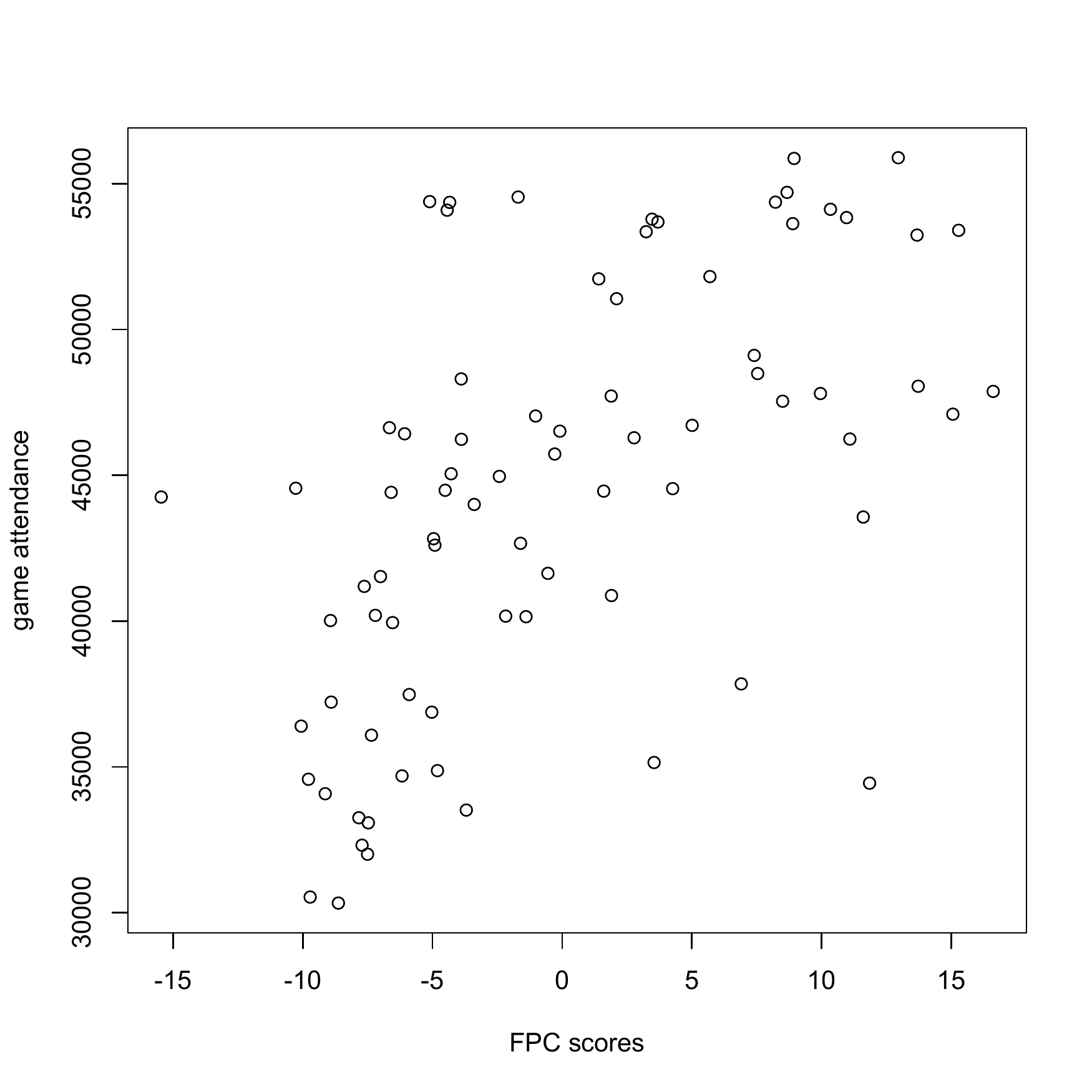}
  }
  \caption{Left: $L^2$ eigenfunctions of $\hat{C}^{+}_\mathsf{trace}$. Right: Scatterplot of game attendance versus functional principal component scores (with respect to the first eigenfunction).}\label{fig:real:eig_score}
\end{figure}

To provide further insights of such phenomenon, we investigate the $L^2$ eigen-decomposition
of $\hat{C}^{+}_\mathsf{trace}$.
Due to the built-in low-rank estimation, $\hat{C}^{+}_\mathsf{trace}$ is automatically of rank $5$ without further truncation of eigenvalues. Its corresponding five $L^2$ eigenfunctions, as described in Section \ref{subsec:eigfun}, are shown in Figure \ref{fig:real:eig_score} (Left).
The first eigenfunction explains over 80\% of the total variance, i.e., the
first eigenvalue is greater than 80\% of the sum of all five eigenvalues. 
Therefore, the first eigenfunction plays a major role in the variation
of the traffic profile.
Of interest is that this eigenfunction possesses two peaks located near times $0$ and $50$, where
the second peak is spanning over the time interval roughly between 30 and 120.
This eigenfunction characterizes the high correlation we have observed between
time $0$ and the time interval between $30$ and $120$.
Since a positive variation along this eigenfunction will add traffic to
these two peaks, this implies that
some audiences may choose to leave shortly after the game or even earlier, while
some others take longer than usual to leave.
As suggested by \citet{Zhang-Wang15}, one possible explanation for this phenomenon is high game attendance.
For games with high attendance, one may choose
to leave earlier than usual to avoid traffic. Meanwhile,
heavy traffic would also last longer due to high attendance.
To further verify this explanation,
we produced the functional principal component (FPC) scores
by pre-smoothing individual vehicle count curves and then projecting them
onto the first eigenfunction.
Smoothing spline with \textsc{gcv} was used to implement the pre-smoothing. 
The scatter plot between FPC scores and game attendance as shown in
Figure \ref{fig:real:eig_score} (Right), together with the fact that their Pearson correlation is $0.57$, 
evidently indicates a positive association.

\linsps
\linsp

\section*{Acknowledgements}
The research of Raymond K.~W.~Wong is partially supported by National Science Foundation grant DMS-1612985. The research of Xiaoke Zhang is partially supported by National Science Foundation grant DMS-1613018.

\bibliographystyle{agsm}
\bibliography{raywongref-all-more}

\end{document}